\documentclass[10pt]{article}

\usepackage{amsmath}
\usepackage{amssymb}
\usepackage{graphics}
\usepackage{rotating}
\usepackage{cite}
\usepackage{color}
\usepackage{fancybox}

\def\0{\mbox{\tiny $0$}}
\def\1{\mbox{\tiny $1$}}
\def\2{\mbox{\tiny $2$}}
\def\3{\mbox{\tiny $3$}}
\def\4{\mbox{\tiny $4$}}
\def\5{\mbox{\tiny $5$}}
\def\6{\mbox{\tiny $6$}}
\def\7{\mbox{\tiny $7$}}
\def\8{\mbox{\tiny $8$}}
\def\9{\mbox{\tiny $9$}}

\def\d{\mbox{\tiny $(4)$}}

\def\M{\mbox{\tiny $M$}}

\renewcommand{\d }[1]{\displaystyle{#1}}
\newcommand{\p }[1]{{\mbox{\tiny #1}}}

\def\spm{\mbox{\tiny SPM}}
\def\ana{\mbox{\tiny ANA}}
\def\num{\mbox{\tiny NUM}}
\title{ \shadowbox{\large \bf RELATIVISTIC TUNNELING THROUGH OPAQUE BARRIERS}}

\author{
\small  Stefano De Leo\thanks{Department of Applied Mathematics,
State University of Campinas, Brazil [deleo@ime.unicamp.br] } \,\,
and\, Vinicius Leonardi\thanks{Department of Applied Mathematics,
State University of Campinas, Brazil}}

\date{\small
\fcolorbox{black}{yellow} {\color{red} $\bullet$ {\color{black}{
{\footnotesize  {\sc Physical Review A} {\bf 83} (2011) 022111-5
}}} {\color{red}{$\bullet$}} } }

\begin{document}
%

\maketitle

\vspace*{-.7cm}

\begin{abstract}
\noindent We propose an analytical study of relativistic tunneling
through opaque bar\-riers. We obtain a closed formula for the
phase time. This formula is in excellent agreement with the
numerical simulations and corrects the standard formula obtained
by the stationary phase method. An important result is found when
the upper limit of the incoming energy distribution coincides with
the upper limit of the tunneling zone. In this case, the phase
time is {\em proportional} to the barrier width.
\end{abstract}
%







\section*{\normalsize I.INTRODUCTION} \label{intro}
The question of how long it takes a particle to tunnel through a
barrier potential has been the issue of intriguing discussions in
the last decades. Since the first studies regarding the tunneling
process \cite{C31,M32}, several tunneling time definitions have
been proposed \cite{H89,O04,W06}. However, there is no general
agreement about a satisfactory definition. Considering a
time-dependent description in terms of wave packets, Hartman
\cite{H62} used the stationary phase method (SPM), previously
applied to scattering problems \cite{B51,W55}, to estimate the
instant in which the transmitted peak appears in the free region
after the barrier. This time definition is known as {\em phase
time}. One can also introduce a time integration over the
probability of finding the particle inside the barrier. This
average time spent in the potential region, regardless whether
transmission or reflection occurs, is known as {\em dwell time}
\cite{B83,L89}. Phase and dwell times have a well established
mutual relation \cite{H89,W03}. The important point to be noted
here is that,  in the opaque limit, both predict the independence
of tunneling times on the barrier width, the so-called Hartman
effect \cite{H62}. It is also possible to investigate transit
times by describing kinematical paths in the potential region
\cite{S87,Iu97,Y99} or introducing a new degree of freedom, like
the Larmor clock \cite{B83} and time-oscillating barriers
\cite{B82}. Recently, a time operator has been considered
\cite{Ba06,O07,Or09}, which is canonically conjugated to the
energy operator. From this time operator it is obtained a
tunneling time that is equal to the average over momentum
components of the phase time.

Although the majority of the discussions about tunneling times
have been based on the non-relativistic Schr\"{o}\-dinger
equation, some recent works have attended to particular features
of relativistic tunneling particles, such as superluminality
\cite{E96,M96,Kr01,N04} and existence of the Hartman effect for
relativistic potentials \cite{W04,L07,DL07,B08b}. The solutions of the Dirac equation are oscillatory in the
diffusion, $E>V_{\p0}+m$ \cite{G99,DL09}, and Klein, $E<V_{\p0}-m$ \cite{K29}, energy zones, where $V_{\p0}$ is the barrier height and $m$ is the particle mass. The phenomenon of multiple peaks occurs in the diffusion zone \cite{DL06b}. The
Klein region, in its turn, involves pair production and
dynamical localized states \cite{DL06}. The tunneling energy zone,
$\max \{m,V_{\p0}-m\}<E<V_{\p0}+m$, is characterized by evanescent
solutions \cite{DL07}. In this paper, we aim to present an
analytic and numerical study of the phase time for one-dimensional
relativistic tunneling in the opaque barrier limit. We choose an incoming spectrum and vary the
barrier height in such a way that the momentum distribution
remains at evanescent zone with above potential energies, i.e.
\begin{equation}
\label{tunz}
 \max\{m,V_{\p0}\}<E<V_{\p0}+m\,\,,
 \end{equation}
whence we observe that energy components with $E<V_{\p0}$ correspond to below potential
states whose interpretation is still subject to discussions \cite{G99,DL06,DL07}.

In section II, we review some of the standard calculations on Dirac tunneling and find an approximation for the transmission coefficient in the opaque limit. This approximation is then used in section III to obtain a closed formula for the phase time in relativistic tunneling. Such an analytic expression corrects the well-known formula obtained by the SPM. The tunneling time is proportional to the barrier width for an incoming
momentum distribution whose upper limit is very close to the barrier
height. This result clearly contradicts the Hartman effect. Our analytical expression and the SPM formula coincide for higher
potentials. To support our analytical study, we also present numerical calculations. Our conclusions are drawn in section IV.

\section*{\normalsize II. OPAQUE BARRIERS AND FILTER EFFECT}
\label{Sec2} Consider a relativistic spin one-half particle of
mass $m$ moving along the $z$-axis in presence of a
one-dimensional electrostatic potential whose height is $V_{\p 0}$
in the region $0<z<L$ and zero elsewhere. The particle dynamics is
described by [$\hbar=c=1$]
\begin{equation}\label{eq Dirac} i\,\partial_t\Psi(z,t)=
\gamma_{\0}\,[\,m-i\,\gamma_{\3}\,\partial_z]\,\Psi(z,t)+V(z)\,\Psi(z,t)
~~,
\end{equation}
where $\gamma_{\0}$ and $\gamma_{\3}$ are the Pauli-Dirac matrices
defined by
$$
\gamma_{\0}=\left(\begin{array}{cc} 1 & 0
\\ 0 & -1
\end{array}\right)~~~~~\mbox{and}~~~~~\gamma_{\3}=\left(\begin{array}{cc} 0 &
\sigma_{\p 3} \\ -\sigma_{\p 3} & 0 \end{array}\right)~~.
$$
Let us consider as incident wave packet
\begin{equation}\label{onda inc} \Psi_{\p{inc}}(z,t)=\d{ N
\int_{p_{_{\p m}}}^{p_{_{\p M}}}dp~g(p)\,u(p)~e^{i\,(p\,z-E\,t)}}
~~,
\end{equation} with $$ g(p)=\exp[-(p-p_{_{\p0}})^{\p2} a^{\p2}/4\,]
~~~~~\mbox{and}~~~~~u(p)=\left(~1~~0~~\dfrac{p}{E+m}~~0~\right)^t~~.
$$ $N$ is a normalization constant and $a$ is related to the  spatial
localization of the incoming particle. In this paper, we choose
\[\mbox{max}\left\{\,0\,,\,\sqrt{V_{\0}^{^{2}}-m^{\2}}\,\right\} \leq p_{\p m}~~~~~
\mbox{and}~~~~~ p_{\p M}\leq \sqrt{V_{\0}^{^{2}}+2\,m\,V_{\0}}
\]
to guarantee that the energy spectrum remains in the tunneling
zone, Eq.\,(\ref{tunz}). This avoids  Klein and diffusion
phenomena, as well the presence of below potential evanescent
solutions.  By solving the Dirac equation (\ref{eq Dirac}) and
imposing the continuity of the wave function at $0$ and $L$, we
find for the transmitted wave packet\cite{G99,DL07,DL09},
\begin{equation}\label{onda tra} \Psi_{\p {tra}}(z,t)=\d{
N~\int_{p_{_{\p m}}}^{p_{_{\p M}}}
dp~g(p)\,T(p,L)\,u(p)~e^{i\,(p\,z-E\,t)}}~~~,
\end{equation}
with
\begin{equation}\label{T}
T(p,L)~=~e^{-ipL}~\left[\cosh(qL)-i~\dfrac{p^{\p2}-EV_{\p0}}{qp}
\sinh(qL)\right]^{^{- 1}}~~, \end{equation}\\ where
$~q=\sqrt{m^{^{\p2}}-(E-V_{\p0})^{^{\p2}}}~$. To summarize our
notation, we shall denote by $q_{_{\p {m,M}}}$ and $E_{_{\p
{m,M}}}$ the $q$ and $E$ functions respectively calculated in
$p_{\p m}$ and $p_{\p M}$.

For thin barriers, the transmitted momentum distribution is very
similar to the incoming distribution. Consequently, the
transmitted momentum is still centered in $p_{\p 0}$. For very
thicker barriers, the mean value of the transmitted momentum,
$\langle p \rangle _{_{\p T}}$, tends to $p_{_{\p M}}$, see
Fig.\,1. By increasing the barrier width, the potential acts as a
momentum filter. Observe that the filter effect is more evident
for momentum distributions whose upper limit coincides with the
upper limit of the tunneling zone, see Fig.\,1b. In the opaque
limit, $mL\gg 1$, we can approximate the expression for $T$ given
in Eq.(\ref{T}) by
\begin{equation}\label{T approx} T(p,L)~\approx ~\dfrac{2}{mV_{\p
0}} ~qp~e^{-qL}~e^{-ip\,L+i\varphi}~~,
\end{equation} where
$\varphi=\arctan\left[(p^{\p2}-EV_{\p0})/qp\right]$.

Before beginning our discussion on the phase time formula obtained
by using the stationary phase method,  let us calculate, in the
opaque limit,  the average momentum
\begin{equation}\label{p medio def} \langle p \rangle _{_{\p
T}}~=~\int_{p_{_{\p m}}}^{p_{_{\p M}}}dp~p~\|g(p)\,T(p,L)\,
u(p)\|^{^{\p2}}\,\,\mbox{\huge /}\,\, \int_{p_{_{\p
m}}}^{p_{_{\p{\p M}}}}dp~\|g(p)\,T(p,L)\, u(p)\|^{^{\p2}}\,\,.
\end{equation}
By using the approximation (\ref{T approx}), we find
\begin{equation}\label{pm2} \langle p \rangle _{_{\p
T}}~\approx~\int_{p_{_{\p m}}}^{p_{_{\p
M}}}dp~p^{\3}q^{\2}~g^{\2}(p)\|\,
u(p)\|^{^{\p2}}e^{-2qL}\,\,\mbox{\huge /}\,\, \int_{p_{_{\p
m}}}^{p_{_{\p{\p M}}}}dp~p^{\2}q^{\2}~g^{\2}(p)\|\,
u(p)\|^{^{\p2}}e^{-2qL}\,\,.
\end{equation}
Observing that $p\,\mbox{d}p=E\,q\,\mbox{d}q\,/\,(V_{\0}-E)$, we
can change the variable of integration from $p$ to $q$,
\begin{equation}\label{pm3} \langle p \rangle _{_{\p
T}}~\approx~\int_{q_{_{\p m}}}^{q_{_{\p
M}}}dq~p^{\2}q^{\3}~\underbrace{[E/(V_{\0}-E)]\,g^{\2}(p)\|\,
u(p)\|^{^{\p2}}}_{f(p)}\,e^{-2qL}\,\,\mbox{\huge /}\,\,
\int_{p_{_{\p m}}}^{p_{_{\p{\p
M}}}}dp~p\,q^{\3}~f(p)\,e^{-2qL}\,\,.
\end{equation}
The filter effect suggests expanding the factors which appear in
the integrands around $q_{\p M}$,
\begin{eqnarray*}
p^{\2}q^{\3} & = & (p^{\2}q^{\3})_{\p M} + (p^{\2}q^{\3})_{q,\p
M}\,(q-q_{\p M}) + \mbox{O}\left[\,(q-q_{\p
M})^{^{2}}\right]\,\,,\\
p\,q^{\3} & = & (p\,q^{\3})_{\p M} + (p\,q^{\3})_{q,\p
M}\,(q-q_{\p M}) + \mbox{O}\left[\,(q-q_{\p
M})^{^{2}}\right]\,\,,\\
f(p) & =  & f_{\p M} + f_{q,\p M}\,(q-q_{\p M}) +
\mbox{O}\left[\,(q-q_{\p M})^{^{2}}\right]\,\,.
\end{eqnarray*}
After algebraic manipulations, we find
\begin{eqnarray}
\label{p medio} \langle p \rangle_{_{\p T}} &\approx &   p_{_{\p
M}} \left\{\,1 + \frac{1}{2\,L}\, \left[
\frac{(p^{\2}q^{\3})_{q,\p M}-p_{\p M}\, (p\,q^{\3})_{q,\p
M}}{(p^{\2}q^{\3})_{\p
M} } \right] \right\}\, \nonumber \\
& = &   p_{_{\p M}} \left\{\,1 + \frac{1}{2\,L}\, \left[
\frac{p_{q,\p M}\,(p\,q^{\3})_{\p M} + p_{\p M}\,
(p\,q^{\3})_{q,\p M}    -p_{\p M}\, (p\,q^{\3})_{q,\p
M}}{(p^{\2}q^{\3})_{\p
M} } \right] \right\}\, \nonumber \\
 &= & p_{_{\p M}}-\dfrac{q_{_{\p M}} E_{_{\p M}}}{2\,p_{_{\p M}}(E_{_{\p
M}}-V_{\p 0})\,L} ~~.
\end{eqnarray}
Thus, for opaque barriers, the momentum distribution is sharply
peaked in the neighborhood of $p_{_{\p M}}$.

The standard phase time formula is obtained by calculating the space-time points in which the phase $\varphi$ is stationary. The maximum
of the wave packet is found by imposing that the derivative of the
phase calculated in  $\langle p \rangle _{_{\p T}}$ is equal to
zero. For opaque barriers,
\[
\left(p\,z - E\,t - p\,L  +\varphi \right)_{p,\p M} =0\,\,.
\]
Consequently at the edge of the barrier, $z=L$, we have $E_{p,\p
M}\,\tau_{\spm}=\varphi_{p,\p M}$. Finally,
\begin{equation}
\label{t spm} \tau_{\spm}=\dfrac{2\,E_{_{\p M}}-V_{\p 0}}{p_{_{\p
M}}q_{_{\p M}}} ~~.
\end{equation}
We observe that, for $E_{\p M}\rightarrow V_{\p 0}+m$,
$\tau_{\spm}\to \infty$. In the next section, we shall overcome
this ambiguity by proposing a different method to calculate the phase
time. It is based on the analytical calculation of the
probability density by mean of the approximation (\ref{T approx}).
We then compare our phase time formula with Eq.(\ref{t spm}) and discuss
in details the validity of the Hartman effect.

 \section*{\normalsize III. REVISING THE PHASE TIME FORMULA} \label{Sec3}
In this section, we propose an analytic method to obtaining the
phase time. This method is essentially based on the search of the
time for which the electronic density is greater at the far edge
of the barrier, $z=L$. This means taking the derivative of the
electronic density with respect to time and finding when it is
equal to zero,
\begin{equation}\label{conditionplot}
\left( \|\Psi(L,t)\|^{^{\p 2}} \right)_{t }=0~~.
\end{equation}
The time solution of the previous equation will be indicates by
$\tau$. Using this time, it is possible to introduce a tunneling
(or transit) velocity, defined by $v_{\p{tun}}=L/\tau$.

Numerical solutions, $\tau_{\num}$, of Eq.(\ref{conditionplot})
were calculated to an incoming gaussian distribution,
$g(p)$, characterized by
\[ ma=10\,\,\,\,\,\mbox{and}\,\,\,\,\,\left(p_{\p m},p_{\0},p_{\p
M}\right)=(0,\mbox{$\frac{1}{2}$},1)\,m\,\,.
\]
For a potential barrier with height $V_{\0}=E_{\p M}-m(\approx
.414\,m)$,  we find that the tunneling velocity tends to a
constant value when we increase the barrier width $L$, see Table 1
and Fig.\,2. We can also see that for increasing potentials, for
example $V_{\0}=0.46\,m(>E_{\p M}-m)$, the transit time tends to a
constant value. This time is in agreement with the phase time
obtained from the stationary phase method, Eq.\,(\ref{t spm}), see
the last row in Table\,1.

Let us now begin our analytical discussion of
Eq.\,(\ref{conditionplot}).  For opaque barriers, we can use the
appro\-xi\-mation given in Eq.\,(\ref{T approx}) and, due to the
filter effect, develop the phase around $k_{\p M}$ (or $q_{\p M}$
if we intend to change the variable of integration from $k$ to
$q$). Using the expansions up to the second order,
\begin{eqnarray}\label{approx fase}
\varphi = \varphi_{\p M} +   \varphi_{q,\p M}\,(q-q_{_{\p M}}) +
\varphi_{qq,\p M}\,(q-q_{_{\p M}})^{^{2}}/\,2 \,+
\mbox{O}\left[\,(q-q_{_{\p M}})^{^{3}}\right]\,\,, \nonumber\\
E = E_{\p M}+  E_{q,\p M}\,(q-q_{_{\p M}}) + E_{qq,\p
M}\,(q-q_{_{\p M}})^{^{2}}/\,2 \, + \mbox{O}\left[\,(q-q_{_{\p
M}})^{^{3}}\right]\,\,,
\end{eqnarray}
and changing the variable of integration into $\rho=q-q_{_{\p
M}}$, we obtain
\begin{equation}\label{int tra}
\Psi_{tra}(L,t)\approx ~
\dfrac{2Ng_{_{\p M}}E_{_{\p M}}u_{_{\p M}}}{mV_{\p0}(E_{_{\p
M}}-V_{\p 0})}~e^{i\varphi_{_{\p M}}-iE_{_{\p M}}t-q_{_{\p
M}}L}~S(t)\,\,,
\end{equation}
with
\begin{equation}\label{int S}
S(t)~:=~\d{ \int_0^{q_{_{\p m}}-q_{_{\p M}}}d\rho~(\rho+q_{_{\p
M}})^{^{2}}~e^{-\rho L}~\exp\left\{\,i\left[\,\left( \varphi_{q,\p
M} - E_{q,\p M}\, t \right)\,\rho + \frac{\varphi_{qq,\p M} -
E_{qq,\p M}\,t }{2}\, \rho^{\p 2} \right]\right\}}\,\,.
\end{equation}
At this point, it is convenient to introduce the quantities
$\alpha$ and $\beta$ defined by
\begin{eqnarray}\label{alpha beta}
\alpha & =& \alpha_{\1} - \alpha_{\2}\,t=  \varphi_{q,\p M} -
E_{q,\p M}\, t = \dfrac{V_{\p 0}-2E_{_{\p M}}}{p_{_{\p M}}(E_{_{\p
M}}-V_{\p 0})} - \dfrac{q_{_{\p M}}}{V_{\p 0}-E_{_{\p M}}}\,t
\,\,,
 \nonumber \\
\beta & =& \beta_{\1} - \beta_{\2}\,t=\varphi_{qq,\p M} - E_{qq,\p
M}\, t = \dfrac{q_{_{\p M}}\,[\,m^{\2}V_{\0}+E_{\p M}(2E_{\p
M}V_{\0}-2\,E^{^{2}}_{\p M} - V_{\0}^{^{2}})\,]}{2\,p^{^{\p3}}_{\p
M} (E_{_{\p M}}-V_{\p 0})^{^{\p3}}}- \dfrac{m^{\p2}}{2\,(V_{\p
0}-E_{_{\p M}})^{^{\p 3}}}\,t\,\,.
\end{eqnarray}
The observation that for $q_{\p M}\to 0$ the main time
contribution comes from the beta term and that for increasing time
this term (which is coupled to $\rho^{\2}$) becomes comparable to
the $\alpha$ term (proportional to $\rho$), suggests to consider
the following approximation for the exponential which appears in
Eq.(\ref{int S}),
\[
\exp\{\,i\,[\,\alpha \rho+ \beta \rho^{\p{\p 2}}\,]\,\}~\simeq
\,1+i\,(\beta \rho^{\p2}+\alpha \rho)- \mbox{$\frac{1}{2}$}\,
\alpha^{\p2}\rho^{\p
2}-\mbox{$\frac{1}{2}$}\,\beta^{\p2}\rho^{\p4}-\alpha\beta
\rho^{\p3}\,\,.
\]
This means that in the calculation of the integral Eq.(\ref{int
S}), we shall find integrals of the form
\begin{equation}\label{sn}
s(n) := \d{ \int_0^{q_{_{\p m}}-q_{_{\p
M}}}d\rho~(\rho^{\p{n+2}}+2\,q_{_{\p
M}}\rho^{\p{n+1}}+q^{\p2}_{_{\p M}}\rho^{\p n})~e^{-\rho L}} ~
\d{\simeq \dfrac{(n+2)!}{
L^{^{^{\p{n+3}}}}}\left[1+\dfrac{2\,q_{_{\p
M}}L}{n+2}+\dfrac{q^{\p 2}_{_{\p M}} L^{^{\p2}}}{(n+2)(n+1)}
\right]}\,\,.
\end{equation}
Finally,
\begin{eqnarray}\label{modulus}
|S(t)|^{^ {2}} & \approx & \left|\, s(0)- \mbox{$\frac{1}{2}$}\,
\alpha^{\p2}\rho^{\p
2}-\mbox{$\frac{1}{2}$}\,\beta^{\p2}\rho^{\p4}-\alpha\beta
\rho^{\p3}\, +i\,(\beta \rho^{\p2}+\alpha \rho) \right|^{^{2}}
\nonumber
\\
 & \approx &
s^{\p2}(0)+\alpha^{\p2}  \left[s^{\p2}(1)-s(0)s(2)\right]
+2\,\alpha\beta \left[s(1)s(2)-s(0)s(3)\right] +\beta^{\p2}
\left[s^{\p2}(2)-s(0)s(4)\right]\,\,.
\end{eqnarray}
Our phase time formula is obtained by taking the derivative
with respect to time and imposing that it is equals to zero, i.e.
\[
\alpha\, \alpha_t\, \left[s^{\p2}(1)-s(0)s(2)\right] +
(\alpha\,\beta_t+\alpha_t\,\beta)\,\left[s(1)s(2)-s(0)s(3)\right]
+\beta\,\beta_t\, \left[s^{\p2}(2)-s(0)s(4)\right]=0\,\,.
\]
Observing that $\beta_{\1}\beta_{\2}\,
\left[s^{\p2}(2)-s(0)s(4)\right]\ll
\alpha_{\1}\alpha_{\2}\,\left[s^{\p2}(1)-s(0)s(2)\right]$, we find
the following analytical expression  for the transit time,
\begin{equation}\label{t geral}
\tau_{\ana} =
\frac{\alpha_{\1}\alpha_{\2}\,\left[s^{\p2}(1)-s(0)s(2)\right] +
(\alpha_1\beta_2+\beta_{\1}\alpha_{\2})\,\left[s(1)s(2)-s(0)s(3)\right]
}{\alpha_{\2}^{\2}\, \left[s^{\p2}(1)-s(0)s(2)\right] + 2\,
\alpha_{\2}\beta_{\2}\,\left[s(1)s(2)-s(0)s(3)\right] +
\beta_{\2}^{^{2}}\,\left[s^{\p2}(2)-s(0)s(4)\right]}  \,\,.
\end{equation}
In the limit $q_{\M}\to 0$ (which implies  $\alpha_{\2}\to 0$),
the previous expression simplifies into
\begin{equation}\label{t qM0}
\tau_{\ana}\to \frac{\alpha_{\1}}{\beta_{\2}}\,\frac{
\left[s(1)s(2)-s(0)s(3)\right] }{
\left[s^{\p2}(2)-s(0)s(4)\right]} =
\frac{2\,(2\,m+V{\0})}{\sqrt{V_{\0}^{^{2}}+2\,mV_{\0}}}\,
\frac{(3!\,4!-2!\,5!)/L^{^{9}}}{(4!\,4!-2!\,6!)/L^{^{10}}}=\frac{2}{9}\,
\sqrt{1+\frac{2\,m}{V_{\0}}}\,L\,\,.
\end{equation}
The SPM result is recovered for $q_{\p M}$ values for which the
time dependence in $\alpha$ is not more negligible. In such a
limit, the $\beta$ term plays no role in the calculation and
\begin{equation}\label{t qM}
\tau_{\ana}\to \frac{\alpha_{\1}}{\alpha_{\2}} = \dfrac{2\,E_{\p
M}-V_{\p 0}}{p_{\p M}\,q_{\p M}} ~~.
\end{equation}
The phase time formula (\ref{t geral}), the formula obtained
by the stationary phase method (\ref{t spm}) and the numerical
simulations are shown in Fig.\,3. Our phase time formula is in
excellent agreement with the numerical analysis.

Before concluding this section, let us make some observations. In
the case, $q_{\M}=0$, the spatial phase $\exp[i\,k(x-L)]$ can be
approximated  by using
\[
k = \sqrt{V_{\0} (V_{\0}+m)} - \frac{V_{\0}+m}{2\,m\,\sqrt{V_{\0}
(V_{\0}+m)}}\,\,q^{\2} \,+\mbox{O}\left[\,q^{^{3}}\right]\,\,.
\]
This implies that the spatial dependence has to be included  in
the $\beta$ term (which contains the  terms in $q^{\2}$). To
reduce the maximum of $S(t)$, see Eq. (\ref{modulus}), of a factor
$1/e$, observing that the time (\ref{t qM0}) guarantees
$\alpha^{\2}\sim\alpha \beta\sim \beta^{\2}$, we have to consider
$x-L\sim \tau_{\ana}$. Consequently, the spreading of the wave
packet in configuration space is {\em proportional} to the barrier
width.

It is also interesting to note that in the non-relativistic
limit, i.e. $E\ll m$ and $V_{\0}\ll m$, the transit velocity
$L/\tau_{\ana}$ tends to $4.5\,\sqrt{V_{\0}/2\,m}$, which is
clearly subluminal.

\section*{\normalsize IV. CONCLUSIONS}
\label{Sec4} In this paper, we have investigated the tunneling
through opaque barriers for relativistic particles. By an
analytical study, we have found a closed formula for the phase
time in agreement with numerical simulations. In order to avoid
``negative energies'' in the potential region, the incoming
momentum spectrum has been restricted to evanescent zone with
$E>V_{\p0}$.  We have assumed that tunneling time $\tau$ is
defined as the instant of maximum probability of finding the
transmitted particle at the barrier edge, that is to say, the peak
of the wave packet along the time for fixed $z=L$. Due to wave
packet spreading, this measurement differs from mapping of the
peak dynamics along the $z$-axis. Nevertheless, numerical
calculations, which can be readily performed, guarantee that such
variation is not relevant. An approximation on the transmission
coefficient has allowed us to obtain a closed expression for
$\tau$. This formula has exhibited excellent agreement with the
numerical calculations and has shown cases in which the stationary
phase method gives a satisfactory approximation. The most
important prediction of our analytical formula concerns the
validity  of the Hartman effect. It does not hold for incoming
distributions whose maximum energy value coincides with the upper
limit of the evanescent zone. In this case, the phase time
increases linearly as function of the barrier width. In this limit
the result obtained by the SPM becomes meaningless since its
formula gives an infinity. For incident wave packets with momentum
integration truncated before $V_{\p 0}+m$, we have found  a
tunneling time which  is independent of $L$ in the opaque limit,
as predicted by the Hartman effect.

Superluminal  transit velocities appear for small barriers, see
Fig.\,3. This phenomenon surely deserves further investigations.
Nevertheless, it has to be underlined that there is {\em no}
causal connection between the peak of the incoming wave packet and
the peak of the transmitted wave packet. One can consider energy
components and barrier heights such that new tunneling features
are introduced, namely negative $E-V_{\p 0}$ values and a Klein
zone that borders the evanescent one.

The numerical simulations presented in these paper are based on
the numerical calculation of the gaussian convolution of the
barrier stationary solutions. Braun, Su and Grobe\cite{BSG} have
proposed a numerical approach, based on the split-operator
technique, to solve the time-dependent three-dimensional Dirac
equation. In view of the results presented in this paper, it could
be interesting to revise tunneling phenomena by using the BSG
approach.

\section*{\small \rm ACKNOWLEDGEMENTS}

The authors thanks the referee for his observations and his very
useful suggestions. One of the authors (SdL) wish also to thank
the Department of Physics, University of Salento (Lecce, Italy),
for the hospitality and the FAPESP (Brazil) for financial support
by the Grant No. 10/02216-2.

\newpage

\vspace*{1cm}

\begin{table}[h!] \begin{center}
\label{Table1} \vspace*{-0.8cm}\hspace*{0cm}
\begin{tabular}{|c|c|c|c|c|} \hline &
\multicolumn{2}{c|}{$V_{\p0}=E_{_{\p M}}-m$}   &
\multicolumn{2}{c|}{$V_{\p0}=0.46~m$}  \\ \hline \ $m\,L$ \ & \ \
$m\,\tau_{\num}$ \ \  & $\tau_{\num}/\tau_{\ana}$ & \  \
$m\,\tau_{\num}$ \ \  & $\tau_{\num}/\tau_{\ana}$ \\ \hline 50 &
23.62 & 0.8807  & 6.520  & 0.9448  \\ \hline 100 & 51.25  & 0.9553
& 7.339  & 0.9906   \\ \hline 150 & 77.87  & 0.9677  & 7.552  &
0.9963 \\ \hline 200 & 104.3  & 0.9719  & 7.650  & 0.9981
\\ \hline 250 & 130.6  & 0.9739  & 7.706  & 0.9988 \\ \hline 300 &
156.9  & 0.9749  & 7.743 & 0.9992   \\ \hline 350 & 183.2  &
0.9755  & 7.769  & 0.9994 \\ \hline 400 & 209.4  & 0.9759  & 7.788
& 0.9996   \\ \hline 450 & 235.7 & 0.9762 & 7.803 & 0.9997 \\
\hline 500 & 261.9 & 0.9764 & 7.815 & 0.9997 \\ \hline \hline
$m\tau_{\p{{\mbox{\sc spm}}}}$ & \multicolumn{2}{c|}{$\infty$} &
\multicolumn{2}{c|}{7.918} \\ \hline \end{tabular}
\caption{Tunneling times are listed for an incoming gaussian
distribution with $ma=10$, $p_{_{\p m}}=0$, $p_{_{\p 0}}=m/2$ and
$p_{_{\p M}}=m$ varying the barrier width  $L$. The barrier height
plays a fundamental role in their characterization. When $V_{\p
0}+m$ is  greater than $E_{_{\p M}}$, for $mL\gg 1$ the tunneling
time tends to a constant value. In this case, the standard formula
obtained by the SPM, Eq. (\ref{t spm}), represents a good
approximation. On the other hand, if $V_{\p 0}+m=E_{_{\p M}}$,
i.e.  $q_{_{\p M}}=0$, the tunneling time is proportional to $L$.
The agreement between analytic, Eq. (\ref{t geral}), and numerical
data is impressive. }
\end{center}
\end{table}

\newpage

\begin{figure}[hbp] \begin{center} \label{Figure1}
\vspace*{-1cm} \hspace*{-2cm} \includegraphics[width=19cm,
height=24cm, angle=0]{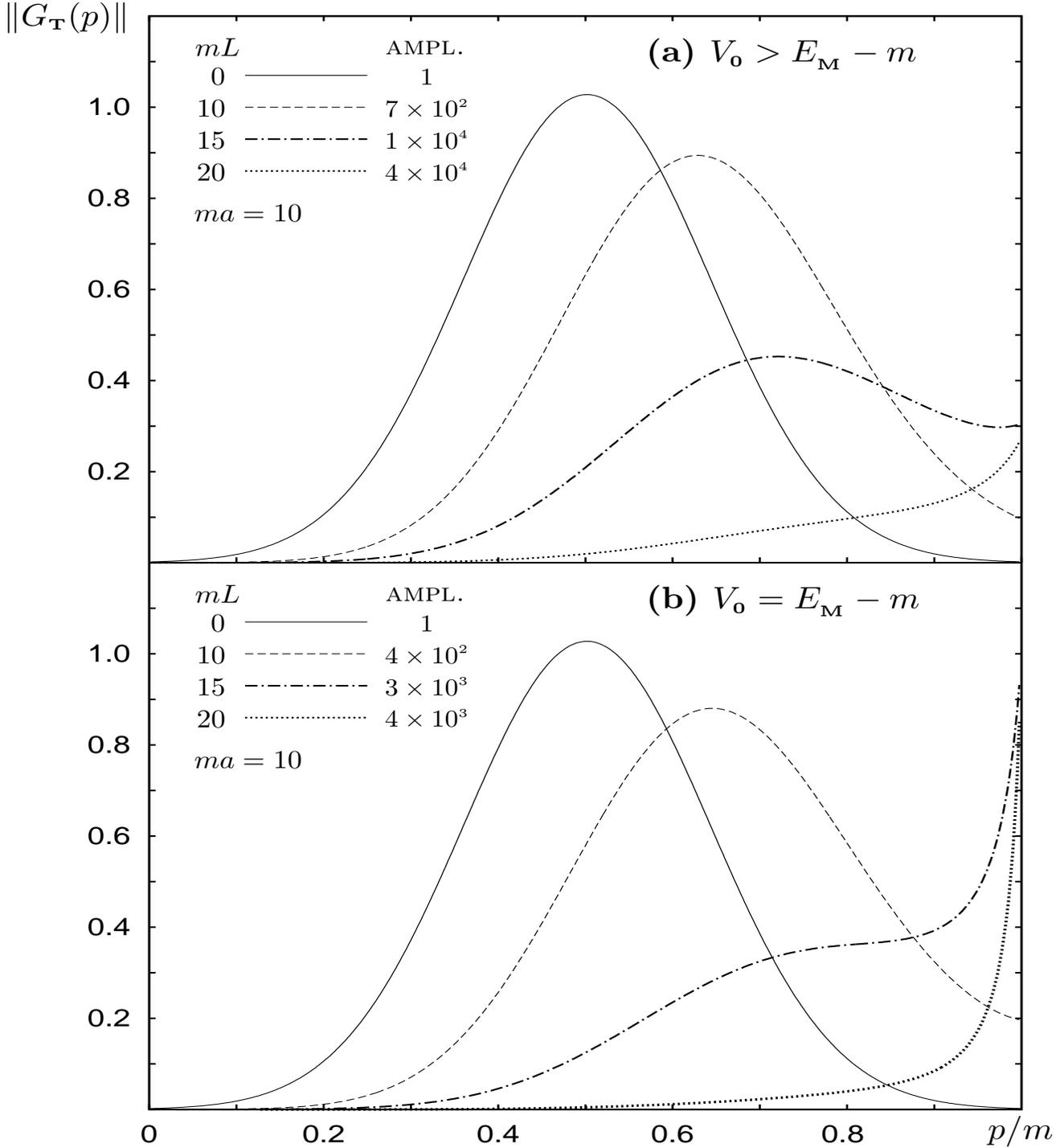} \vspace*{-3cm}
\caption{Transmitted momentum distributions, $G_{\p
T}(p)=g(p)T(p,L)u(p)$, are plotted for different barrier widths.
Continuous line represents the incoming gaussian spectrum with
$p_{_{\p m}}=0$, $p_{_{\p 0}}=m/2$, $p_{_{\p M}}=m$ and $ma-10$.
The barrier heights are chosen equal to $V_{\p 0}=E_{_{\p
M}}-m\approx 0.414\,m$ in (b) and $V_{\p 0}=0.46m$ in (a). Filter
effect is evident in both cases.} \end{center} \end{figure}

\newpage

\begin{figure}[hbp] \begin{center} \label{Figure2} \vspace*{-1cm}
\hspace*{-2cm} \includegraphics[width=19cm, height=24cm,
angle=0]{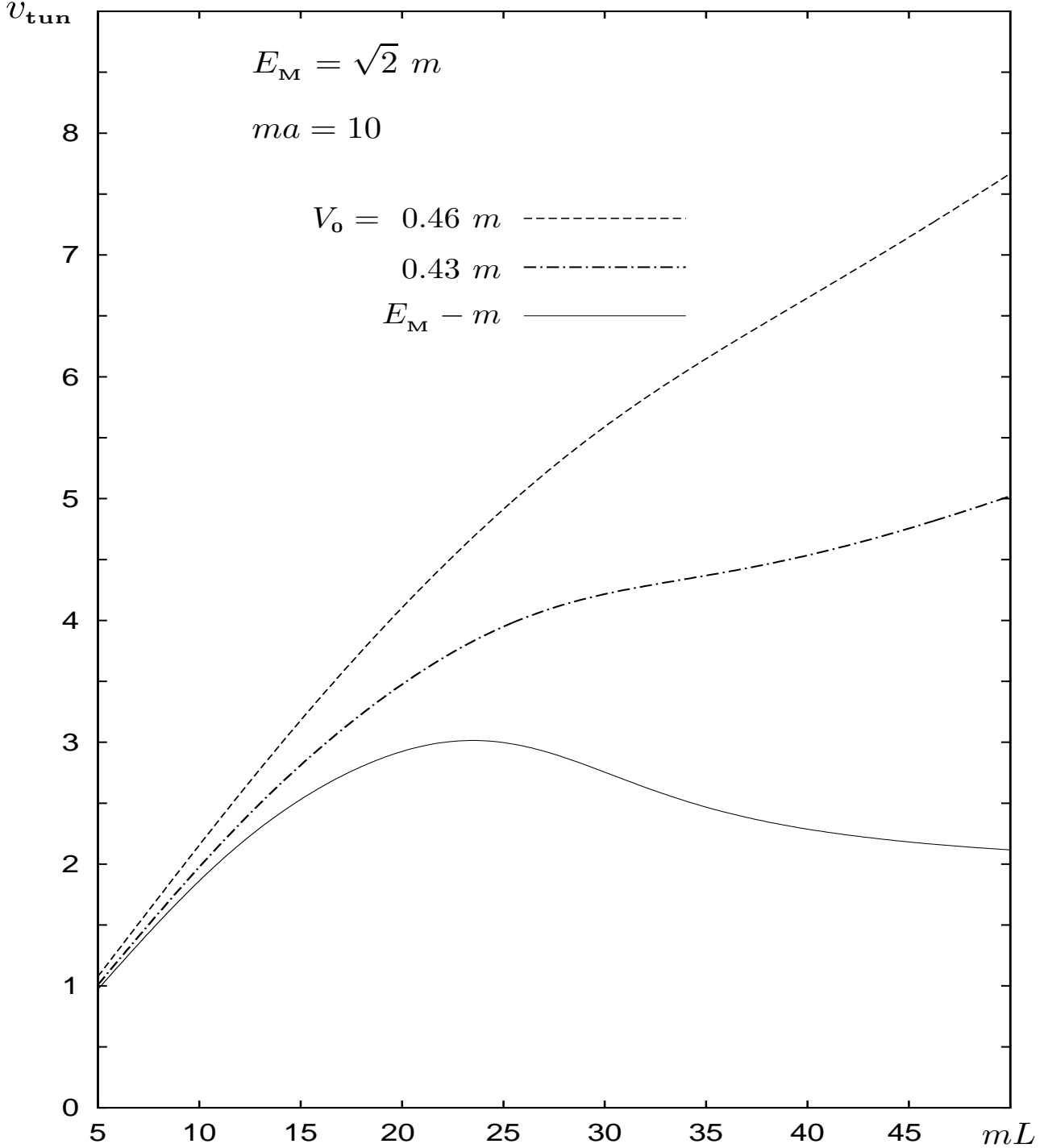} \vspace*{-3.cm} \caption{Tunneling velocities
are plotted as function of $mL$. They have been numerically
obtained by the expression (\ref{conditionplot}), with $p_{_{\p
m}}=0$, $p_{_{\p 0}}=m/2$, $p_{_{\p M}}=m$ and $ma=10$. Actually,
if $V_{\p 0}+m$ is greater than $E_{_{\p M}}$, then $v_{\p{tun}}$
increases linearly with the barrier width (Hartman effect).
However, if the incoming energy distributions reaches the upper
tunneling zone, $E_{_{\p M}}=V_{\p 0}+m$, the velocity tends to a
constant value. Numerical tunneling times for $L>50/m$ are given
in Table 1.} \end{center} \end{figure}

\newpage

\begin{figure}[hbp] \begin{center} \label{Figure3} \vspace*{-1cm}
\hspace*{-2cm} \includegraphics[width=19cm, height=24cm,
angle=0]{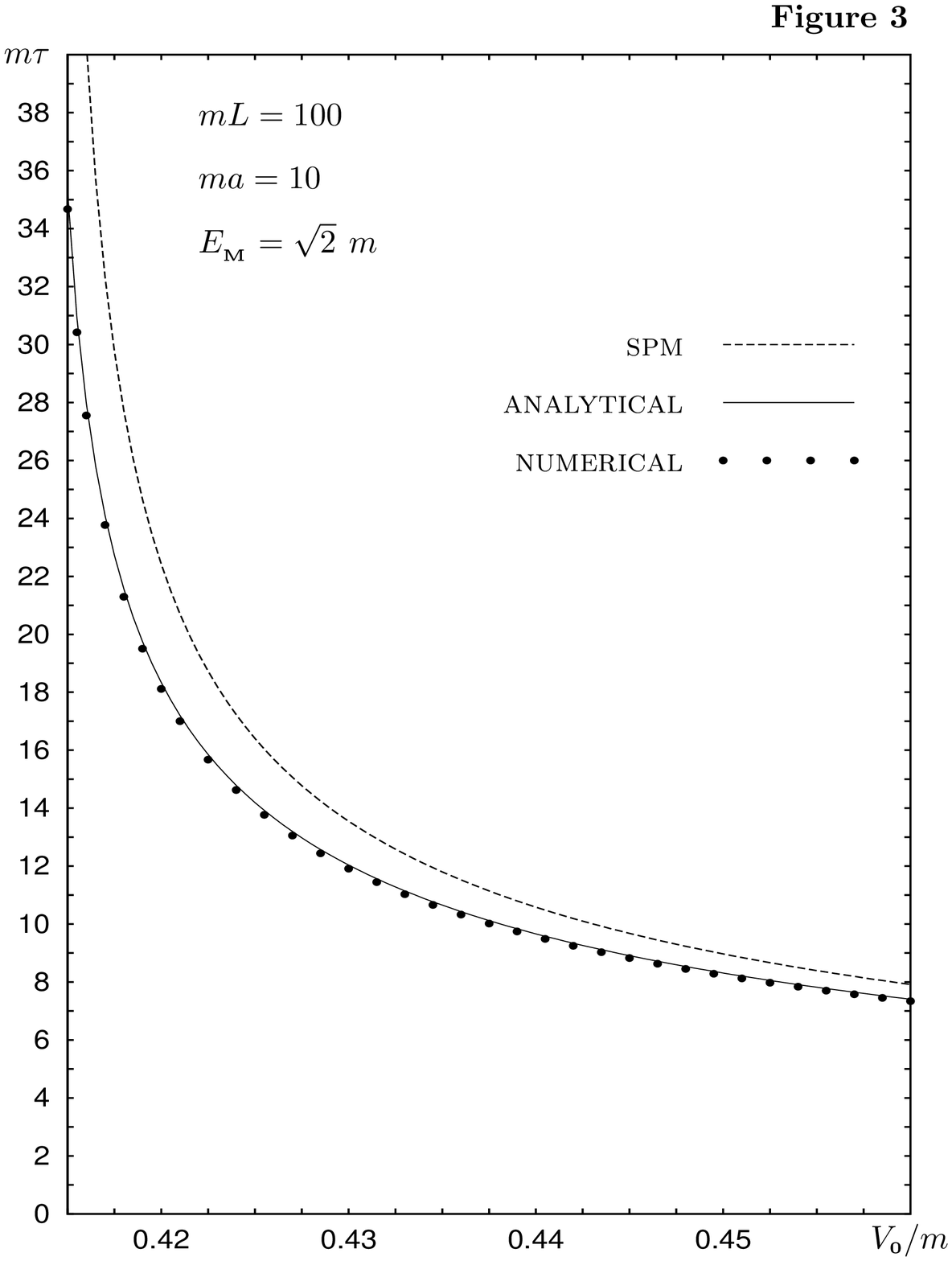} \vspace*{-3cm} \caption{Analytical (ANA),
numerical (NUM) and SPM tunneling times are plotted as function of
$V_{\p 0}/m$. The incoming momentum spectrum is characterized by
$p_{_{\p m}}=0$, $p_{_{\p 0}}=m/2$, $p_{_{\p M}}=m$ and $ma=10$.
The agreement between $\tau_{\ana}$ and $\tau_{\num}$ is
excellent. The SPM represents a good approximation for higher
barriers.}
\end{center} \end{figure} 
\begin{thebibliography}{b!}
\bibitem{C31} E. U. Condon, Rev. Mod. Phys. {\bf 3}, 43 (1931)
\bibitem{M32} L. A. MacColl, Phys. Rev. {\bf 40}, 621 (1932)
\bibitem{H89} E. H. Hauge and J. A. Stovneng, Rev. Mod. Phys. {\bf
61}, 917 (1989)
\bibitem{O04} V. S. Olkhovsky, E. Recami and J.
Jakiel, Phys. Rep. {\bf 398}, 133 (2004)
\bibitem{W06} H. Winful,
Phys. Rep. {\bf 436}, 1 (2006)
\bibitem{H62} T. E. Hartman, J.
Appl. Phys. {\bf 33}, 3427 (1962)
\bibitem{B51} D. Bohm, {\it
Quantum Theory} (Prentice-Hall, New York, 1951)
\bibitem{W55} E.
Wigner, Phys. Rev. {\bf 98}, 145 (1955)
\bibitem{B83} M. B\"{u}ttiker,
Phys. Rev. B {\bf 27}, 6178 (1983)
\bibitem{L89} C. R. Leavens and
G. C. Aers, Phys. Rev. B {\bf 39}, 1202 (1989)
\bibitem{W03} H.
Winful, Phys. Rev. Lett. {\bf 91}, 2604011 (2003)
\bibitem{S87} D.
Sokolovski and L. Baskin, Phys. Rev. A {\bf 36}, 4604 (1987)
\bibitem{Iu97} K. Imafuku, I. Ohba and Y. Yamanaka, Phys. Rev. A
{\bf 56}, 1142 (1997)
\bibitem{Y99} N. Yamada, Phys. Rev. Lett.
{\bf 83}, 3350 (1999)
\bibitem{B82} M. B\"{u}ttiker, R. Laundauer,
Phys. Rev. Lett. {\bf 49}, 1739 (1982)
\bibitem{Ba06} O. del
Barco, M. Ortuño and V. Gasparian, Phys. Rev. A {\bf 74}, 032104
(2006) \bibitem{O07} V. S. Olkhovsky and E. Recami, Int. J. Mod.
Phys. A {\bf 22}, 5063 (2007)
\bibitem{Or09} G. Ordonez and N.
Hatano, Phys. Rev. A {\bf 79}, 042102 (2009)
\bibitem{Kr01} P.
Krekora, Q. Su, R. Grobe, Phys. Rev. A {\bf 64}, 022105 (2001)
\bibitem{E96} T. Emig, Phys. Rev. E {\bf 54}, 5780 (1996)
\bibitem{M96} M. S. Marinov and Bilha Segev, Found. Phys. {\bf
27}, 113 (1996)
\bibitem{N04} G. Nimtz, Found. Phys. {\bf 34},
1889 (2004)
\bibitem{W04} H. Winful, Phys. Rev. A {\bf 70}, 052112
(2004)
\bibitem{L07} J. T. Lunardi and L. A. Manzoni, Phys. Rev. A
{\bf 76}, 042111 (2007)
\bibitem{B08b} A. E. Bernardini, Eur.
Phys. J. C {\bf 55}, 125 (2008)
\bibitem{DL07} S. De Leo, P. Rotelli, Eur. Phys. J. C {\bf 51}, 241 (2007)
\bibitem{G99} F. Gross, {\it Relativistic Quantum Mechanics and Field Theory} (John
Wiley and Sons, New York, 1999)
\bibitem{DL09} S. De Leo and P. P.
Rotelli, Eur. Phys. J. C {\bf 63}, 157 (2009)
\bibitem{K29} O.
Klein, Z. Phys. {\bf 53}, 157 (1929)
\bibitem{DL06b} S. De Leo, P.
Rotelli, Eur. Phys. J. C {\bf 46}, 551 (2006) \bibitem{DL06} S. De
Leo, P. Rotelli, Phys. Rev. A {\bf 73}, 042107 (2006)
\bibitem{BSG}
J. W. Braun, Q. Su, and R. Grobe, Phys. Rev. A {\bf 59}, 604
(1998).
\end{thebibliography}
\end{document}